\documentstyle[11pt,mrs2001,epsfig]{article}
\begin{document}
\title{Tilted CDM versus WDM in the Subgalactic Scuffle}
%

\author{J. Bullock$^1$}
\affil{$^1$Department of Astronomy and Department of Physics, 
The Ohio State University, Columbus, OH
43210; james@astronomy.ohio-state.edu}

\begin{abstract}
Although the  currently  favored cold   dark matter plus  cosmological
constant model (LCDM) has proven to  be remarkably successful on large
scales, on subgalactic    scales it   faces some potentially     fatal
difficulties;  these include  over-producing  dwarf satellite galaxies
and predicting excessive  central densities in  dark halos.  Among the
most natural  cosmological solutions to these  problems is  to replace
cold dark matter with a warm species  (LWDM).  The warm component acts
to reduce the small-scale power, resulting  in fewer galactic subhalos
and lower  halo central densities.  An alternative   model with a mild
``tilt'' in  the inflationary  power   spectrum  (TLCDM;  $n =   0.9$)
similarly reduces the central  densities  of dark halos, although  the
substructure abundance remains  relatively  high.  Here I   argue that
because dwarf galaxy formation should be suppressed in the presence of
a  strong  ionizing background,  favored  LWDM  models will  generally
{\textit{under}-}predict the   observed abundance of   dwarf galaxies.
The satellite count  for   TLCDM fairs much   better,  as long as   the
photoionization effect is taken  into account.  TLCDM provides a  more
successful  alternative to LWDM  on subgalactic scales  with the added
attraction that it  relies on only a minor,  natural adjustment to the
standard framework of CDM.
\end{abstract}

\section{The Matchup}
Inflation-generated cold dark  matter (CDM) models provide an  elegant
and well-motivated  class of theories for the   origin of structure in
the universe.   A    diverse   set    of  large-scale    observational
measurements, both  at high and low  redshifts, are well-accounted for
within the CDM picture, and most point to a single model with $\Omega_m
\simeq 0.3$, $h \simeq 0.7$, and with space made flat by a significant
cosmological                         constant                component
(LCDM) ~\cite{cmb,sn,peacock}.  

On the scales  of  galaxies and  smaller,  the situation is  much more
uncomfortable.  High-resolution, dissipationless N-body simulations have
have led to robust predictions for  the distribution of dark matter in
and   around     (L)CDM    galactic halos~\cite{profiles,dwarfs}.   The
predictions have allowed   detailed comparisons with  observations  on
subgalactic scales and have  lead to two  major discrepancies.  First,
the number  of predicted subhalos with  circular velocities  $\sim
10-30 {\rm ~km~s^{-1}}$ within the virialized extent of Milky Way-type
halos is roughly   an  order of  magnitude  higher than  the  observed
abundance   similarly-sized satellites~\cite{dwarfs}.  Perhaps      the
biggest problem concerns the  central densities of galaxies.  Rotation
curves of low-mass, dark-matter   dominated galaxies seem to  indicate
lower  central densities  than would be   expected ~\cite{dmdom} in the
standard  LCDM model.  A similar problem  may be present in bright
galaxies  ~\cite{bright,ens}, including  our   own Milky Way  
~\cite{mwprobs,ens} (although see ~\cite{vdb,klypin}).

These discrepancies  have   motivated the exploration   of alternative
scenarios.   By stripping either  the collisionless or cold properties
of traditional CDM, or by considering additional exotic possibilities,
many   authors have sought  to  preserve the success   of CDM on large
scales while modifying the manifestations of CDM on small scales
~\cite{altcdm}.  Among  these possibilities,    a popular choice  is  to
replace CDM  with warm dark  mattter  (WDM), thereby suppressing power
below  the free-streeming scale of  the warm 
particle\footnote{$R_f \simeq 0.1 [m_w/{\rm  KeV}]^{-4/3}$Mpc, for 
$\Omega_w = 0.3, h=0.7$, where $m_w$ is the mass of the WDM particle.}
~\cite{wdm,ens}.  A  less  radical, and  perhaps more  naturally  motivated
mechanism  for suppressing the  small scale  power  is to tilt the
initial inflationary   power spectrum to   favor large  scales (TLCDM,
n=0.9)~\cite{alam}.  In  what follows, I pit  LWDM against TLCDM  in a
bout  to determine which  model can best   match the subgalactic data.
The TLCDM model parameters are those favored by recent Ly$\alpha$ forest
measurements~\cite{la}: 
$n=0.9$, $\Omega_m = 0.4$, $h=0.65$, $\Omega_{\rm L}=0.6$,
and $\sigma_8 = 0.66$. The LWDM model has  
$n=1.0$, $\Omega_m = 0.3$, $h=0.7$, $\Omega_{\rm L}=0.7$, and 
$\sigma_8 = 1.0$, with a $1$ KeV WDM particle.

\begin{figure}
\plottwo{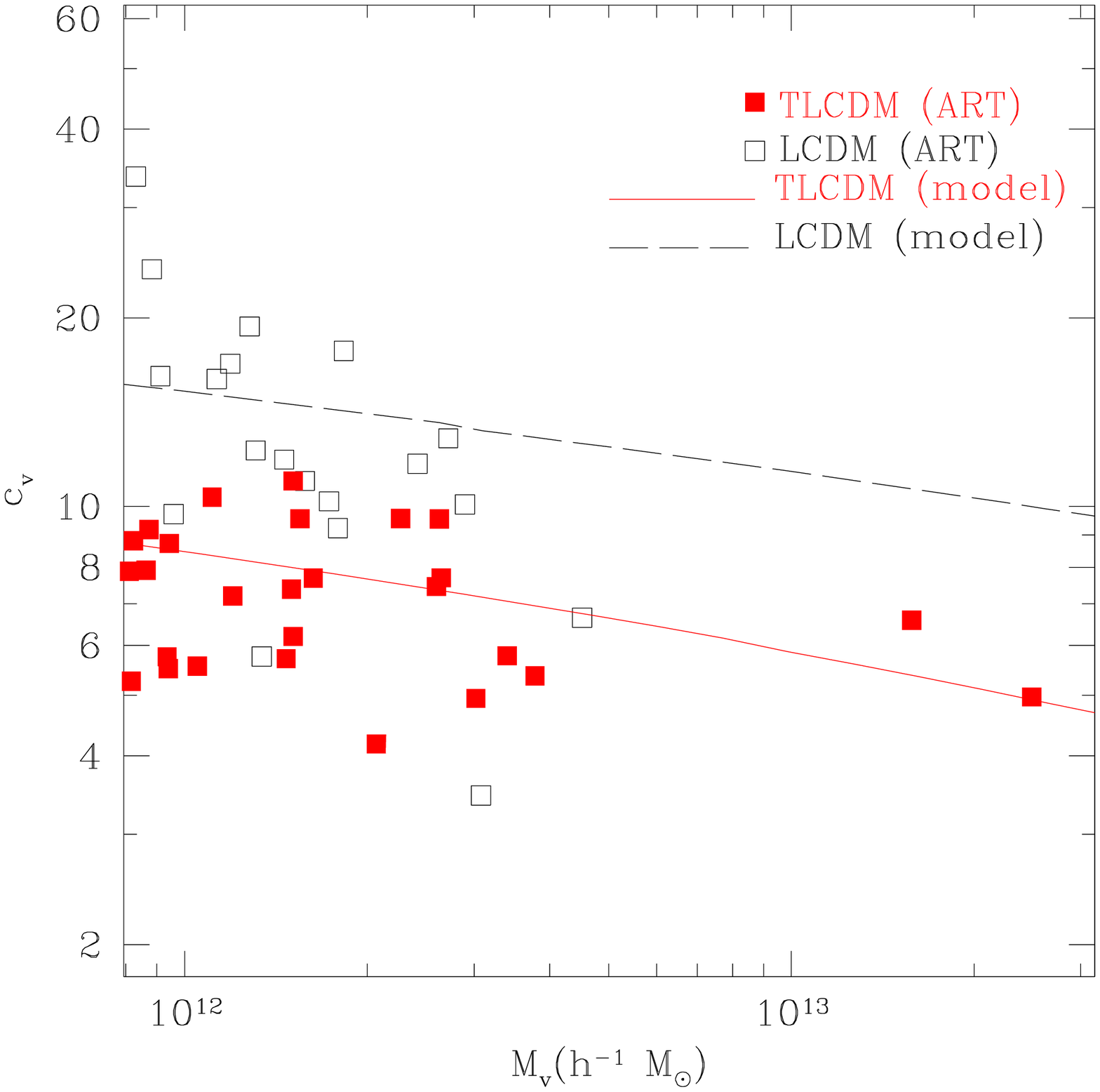}{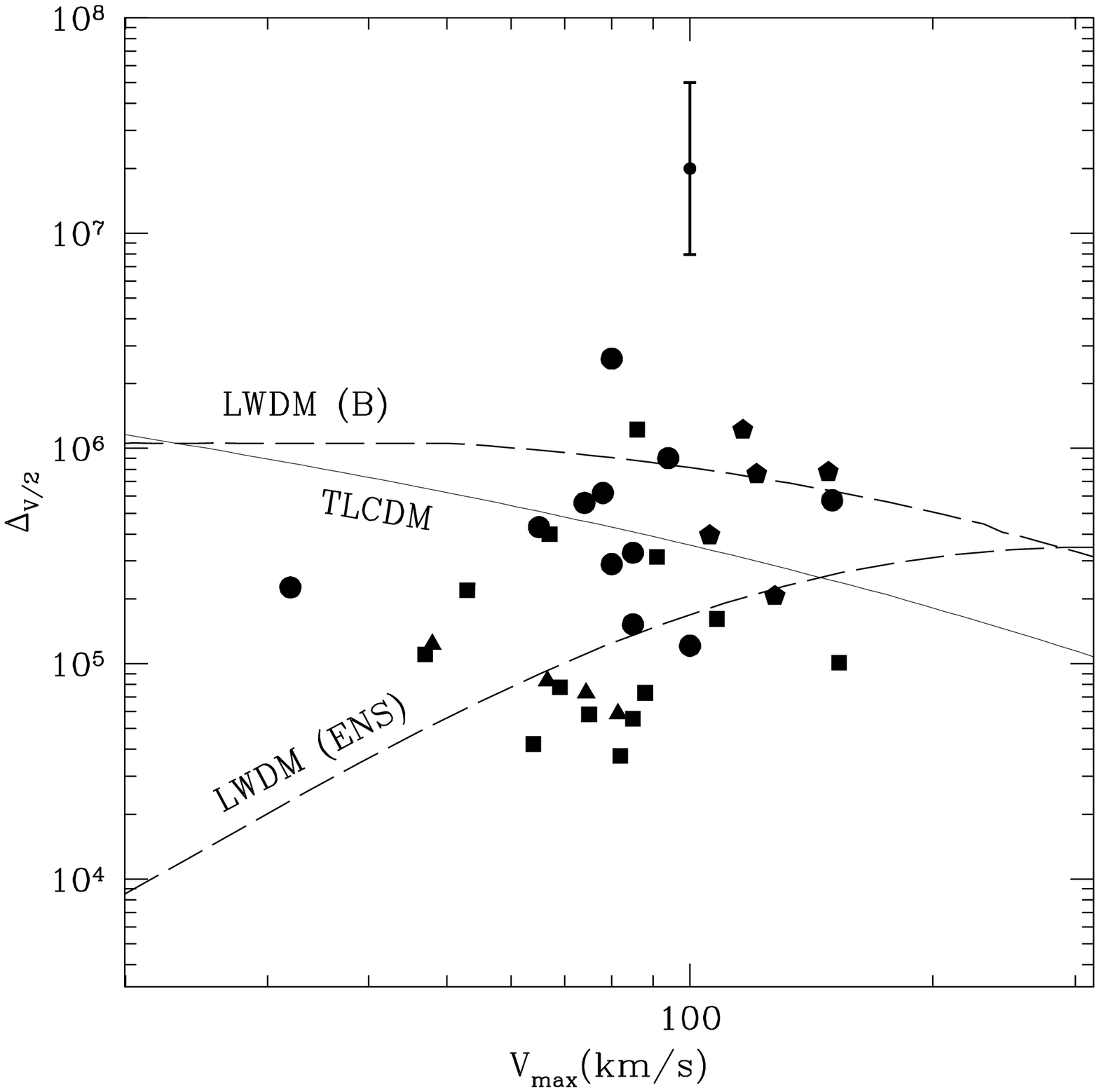}
\caption{(Left) Simulation results for TLCDM and LCDM halo concentrations
as a function of virial mass.  (Right) Central densities of several dark matter
dominated dwarf and LSB galaxies compared with LWDM and TLCDM model
predictions.  The error bar shows the expected scatter in central densities
as reported in references~\cite{bullock,wech}.}
\end{figure}

\section{Round 1: Central Densities}


Halo central densities    are often characterized by  a  concentration
parameter defined as the  ratio of an  outer halo  radius to an  inner
radius where the log-slope of the density  profile is $-2$: $c_{\rm v}
\equiv R_{\rm v}/R_{-2}$.  The outer  radius $R_{\rm v}$ is defined as
the radius within  which the mean  density of  the halo  is a constant
multiple of the critical density (in this case 100 at $z=0$, see, 
e.g.\cite{bullock}   for   details).   A     halo's  central  density    (or
concentration) is set by the density of the  universe at the time when
the  halo's mass accretion rate was  high  ~\cite{wech}, so halos that
form  early are  more   concentrated.  LWDM has significantly  reduced
power on small  scales, therefore low-mass  WDM  halos form  later than
their CDM  counterparts,  and end  up  with  lower $c_{\rm v}$  values
~\cite{wdm}.

A   similar  reduction   in  halo
concentrations occurs in the TLCDM model because the tilt in the power 
spectrum reduces small scale power relative to large.  Figure
1a shows  concentrations  of LCDM  and   TLCDM halos  simulated  
using the  Adaptive   Refinement Tree  Code~\cite{kkk}.   
The  two
simulations consisted  of   boxes of $15 h^{-1}$Mpc   on  a side  with
$128^3$      particles   of mass  $m_p      \simeq   4  \times  10^{8}
h^{-1}$M$_{\odot}$, and they each had an effective  force resultion of 
$4 h^{-1}$kpc.  Solid points represent TLCDM halos, open points represent
LCDM halos, and  the lines show predictions  from the model of Bullock
et al. ~\cite{bullock} (hereafter B) for each of the cosmologies.  At a
fixed mass, the TLCDM halos are roughly  half as concentrated as their
LCDM counterparts.  

Figure 1b illustrates how predicted TLCDM central densities compare to
those   of several dark  matter  dominated Low  Surface Brightness 
(LSB) and
dwarf   galaxies (see ~\cite{alam}   for  a description  of the  data).
Central densities are characterized by  the quantity $\Delta_{\rm V/2}
\equiv \rho(r_{V/2})/\rho_{\rm crit}$.  Here,  $r_{V/2}$ is the radius
where the the rotation velocity  falls to one   half its maximum  value
$V_{\rm max}$,  $\rho(r_{V/2})$  is the   average density within  that
radius,  and $\rho_{\rm crit}$    is the critical density.   The  line
labeled TLCDM  shows  the  B model  prediction  for  TLCDM halos  as a
function of $V_{\rm  max}$.  For  comparison,  the lines  labeled LWDM
show two  different analytic model predictions  for  LWDM halos.
The first  line (ENS)  ~\cite{ens}  follows from  a model developed to
match N-body  simulation  results.  Unfortunately,  the  ENS  analytic
model has not been tested against  N-body simulations over the $V_{\rm
max}$ range that is most relevant for the data shown in Figure 1b. One
may conservatively  expect the properties of  WDM halos  to lie within
the range given by the lines labeled ENS and B.

\section{Round 2: Dwarf Satellites}

One likely solution to  the substructure crisis involves the  expected
supression of low-mass  galaxy formation in  the presence of  a strong
ionizing background  ~\cite{supress}.   In ~\cite{bkw} (BKW)  we pointed
out that the observed satellite abundance is well-accounted for if the
dwarf    galaxy host halos   correspond   only to  those   that had  a
significant fraction of their mass, $f$,  in place before the epoch of
reionization, $z_{\rm  re}$.   Subsequently, this  idea has  been made
more   convincing    by  detailed   semi-analytic    and  hydrodynamic
modeling~\cite{gnedin,somerville,benson}.  Not   only does this  effect
provide a reasonable    solution,  it  seems  to   be   an  inevitable
consequence  of having an ionized  universe.  The implication is that
any model  that makes predictions  involving  dwarf-sized systems must
include photoionization suppression.

Figure 2 illustrates the expected number  of subhalos as a function of
circular velocity 
for a typical  TLCDM Milky-Way halo.  The calculations here were
done following the techniques  outlined in BKW.  For reasonable values
of $z_{re}$ and $f$, the expected  number of observable dwarf galaxies
(shaded region)  is in   good    agreement with the   observed   dwarf
population (points with error bars).  There is similar agreement for a
range of viable  parameter choices.  For  example, the choice  $z_{re}
=6$ and $f=0.25$  works, as does $z_{re} =8$  and $f=0.15$.   The LWDM
model, on the other hand, is unable  to produce any observable dwarfs,
even  if  the reionization  epoch  occurs  at  an unrealistically  low
redshift $z_{re}=5.5$.  Structure forms  much later in the LWDM model,
so  fewer objects collapse  early.  This,   coupled  with the  already
reduced  number of subhalos compared  to  LCDM, makes the abundance of
dwarfs hard to match with a WDM model.  One way out might be if dwarfs
are  produced via the fragmentation  of  larger objects, although this
seems unlikely because mass-to-light ratios of dwarfs suggest
a high dark matter content~\cite{mateo}.

\begin{figure}
\plottwo{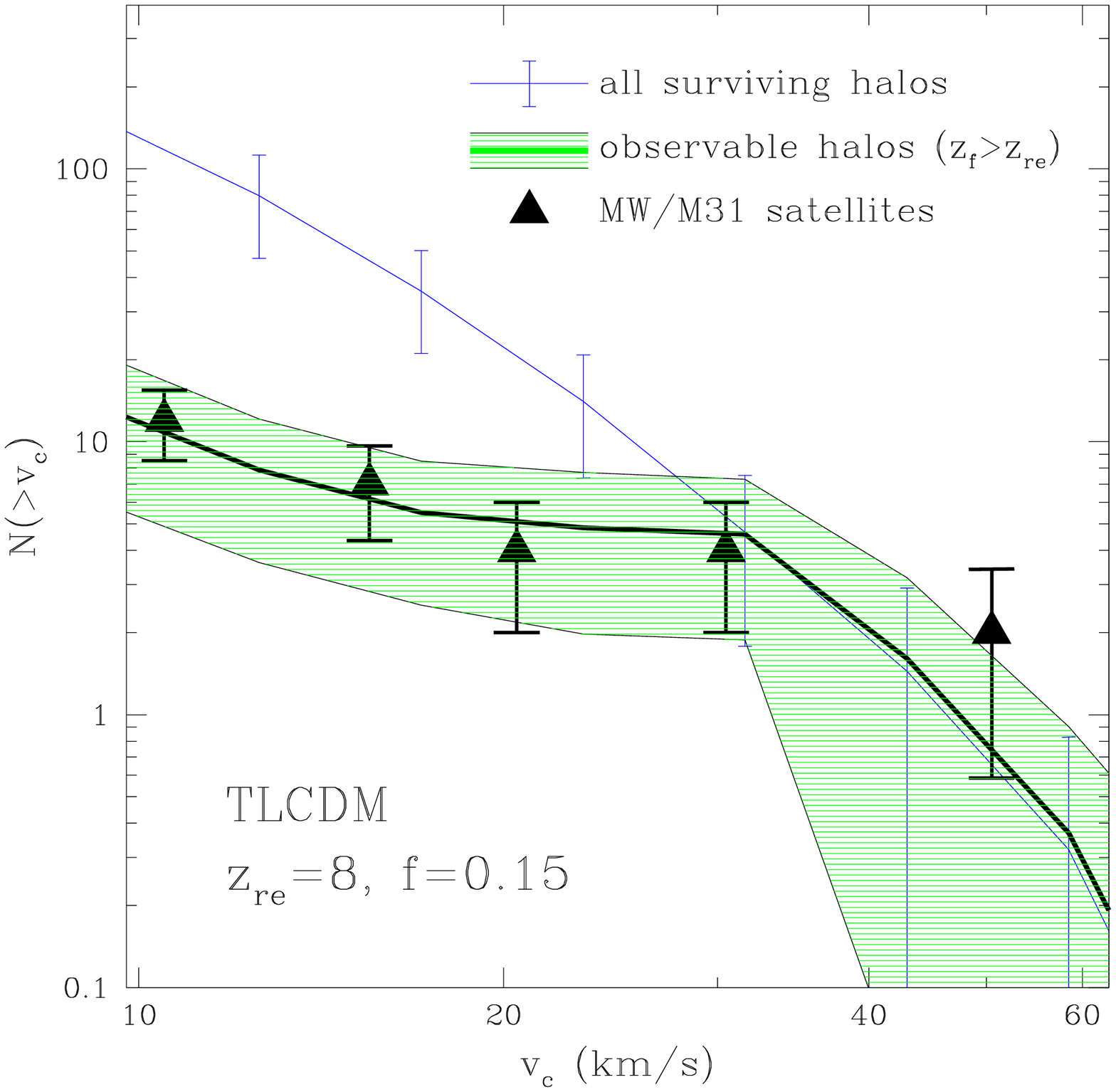}{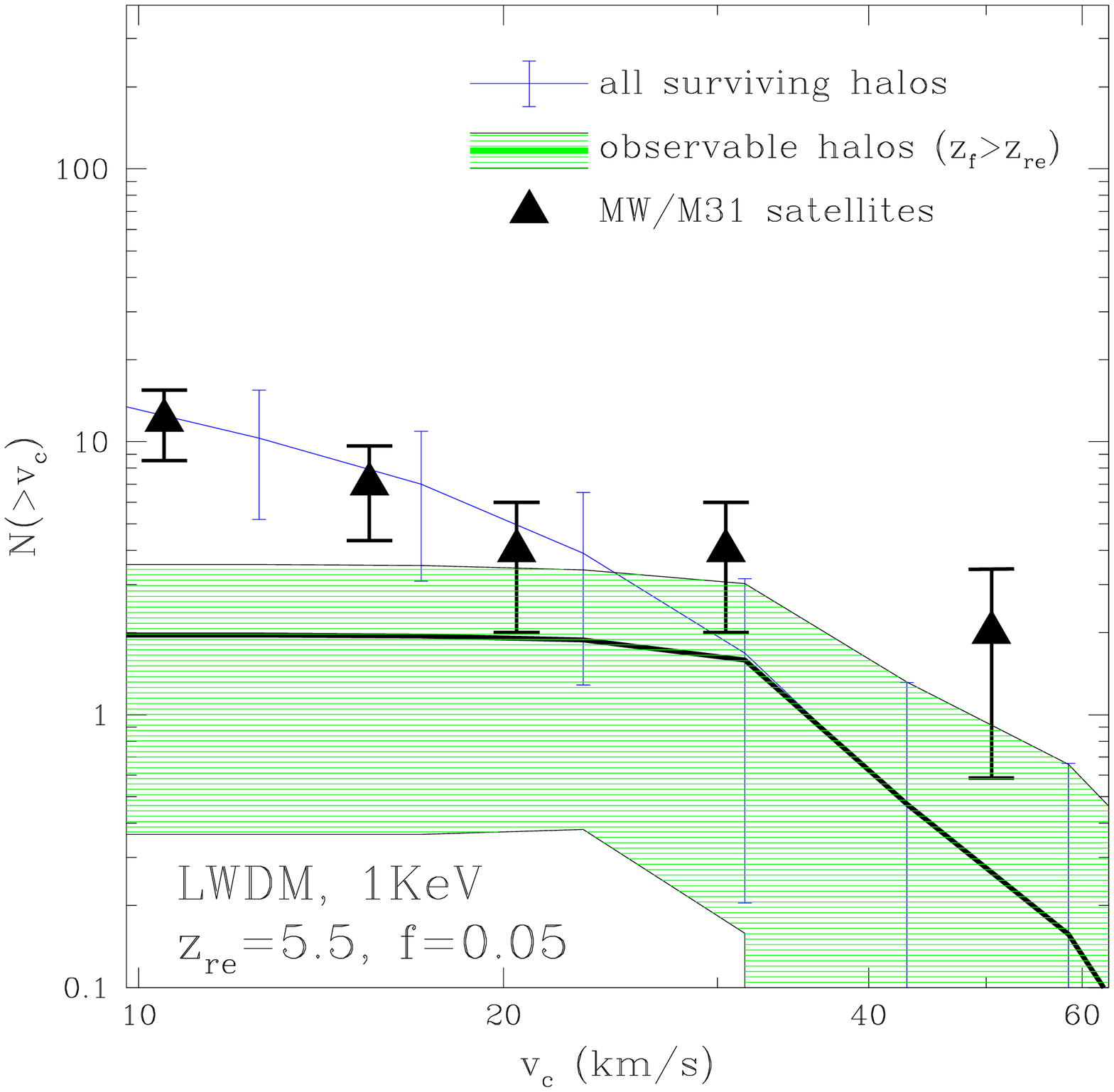}
\caption{(Left) The subhalo (thin line) and observable satellites abundance
(shaded region) for an average 
Milky-Way halo in TLCDM.  Points with error bars
are represent the local group satellites. (Right) The same, except
now for LWDM.}
\end{figure}

\section{The Decision}

WDM helps  the
central density issue by  suppressing  power below the  free streaming
scale of the warm particle.  Halos less massive than a multiple of the
free streaming mass will form later than  their LCDM counterparts, and
therefore have reduced central densities.   For  example, a $1$KeV  WDM
particle will reduce concentrations in  halos with circular velocities
smaller than $\sim 200$km s$^{-1}$.  Figure 1b illustrates how an LWDM
model of this kind provides a reasonable  match to the average central
densities of low-mass dark  matter  dominated systems.  Note that  the
densities  of bright,  Milky-Way-type   galaxies will not  be affected
unless the  particle mass is made smaller  than the $\sim 1$ KeV bound
set by the Lyman-alpha forest ~\cite{nara}.

A TLCDM model normalized  to COBE on  large scales and with parameters
motivated by recent Ly$\alpha$ forest  data provides a similar  remedy.
Instead of a  sharp cutoff at   a free streaming  scale,  the power is
reduced gradually via a long tilted ``lever arm'' ($n=0.9$ rather than
$n=1.0$ in the  primordial spectrum).  Because  the reduction in power
is continuous, all   mass scales collapse   later than  they  would in
standard  LCDM, and the   central  densities in  galaxy-mass halos  are
reduced by roughly a factor of $\sim 4$.  As illustrated in Figure 1b,
TLCDM  does as well   as LWDM in  matching   the central  densities of
a sample dark-matter dominated dwarf and LSB galaxies.
In the  arena of  density comparisons, the   matchup between TLCDM and
LWDM must   be declared a   draw.   
Neither of the  models  solves the
problem decidedly, but they  both  certainly  help relative  to  LCDM.
TLCDM may be   more
desirable   because it is expected   to   reduce densities in  bright
galaxies as well less massive objects.

Compared  to  the  central   density issue,  the   Galactic  satellite
abundance  is much   less troublesome  for  LCDM.   It has  long  been
expected that photoionization (if not super nova explosions ~\cite{ds})
should play a significant role in this mass regime~\cite{supress,bkw}.  
If reionization
is taken into consideration, TLCDM can  account for the observed number
of dwarfs.   LWDM, on the other  hand, faces severe problems.  Without
an ionizing background the   number  of expected subhalos  is  roughly
consistent with the observed  dwarf abundance, but when the background
is included, the number of satellites is vastly under-predicted in LWDM.

Based on two rounds of evidence, TLCDM  is the clear winner over LWDM.
Indeed, the seriousness of the  small-scale density crisis makes TLCDM
a viable  contender for our  cosmology of choice,  a title held now by
standard LCDM.  The  tilt required falls   naturally within the  range
expected in standard inflation models, so  TLCDM is just as attractive
as LCDM in the  ``naturalness''  category.  Forthcoming observational
constraints derived from CMB
studies,   Ly$\alpha$   forest   measurements,  large scale   clustering
analyses, and cluster  counts will provide useful  tests for 
TLCDM.

\acknowledgements{Thanks to Khairul Alam, Andrey Kravtsov, and David
Weinberg for allowing me to present our results here, and to Jeremy
Tinker for comments on a draft.
I am grateful to NASA LTSA grant NAG5-3525 and NSF
grant AST-9802568; without their unwavering support the writing of
this manuscript would have not been possible.}

\vfill
\end{document}